\documentclass[11pt]{article}
\usepackage{moriond,epsfig}

\def\be{\begin{equation}}
\def\ee{\end{equation}}
\def\bea{\begin{eqnarray}}
\def\eea{\end{eqnarray}}

\def\beq{\begin{equation}}
\def\eeq{\end{equation}}
\def\bea{\begin{eqnarray}}
\def\eea{\end{eqnarray}}

\newcommand{\rb}{\mbox{{\bf
r}}}

\def\lsim{\mathrel{\rlap{\lower4pt\hbox{\hskip1pt$\sim$}}
    \raise1pt\hbox{$<$}}}         
\def\gsim{\mathrel{\rlap{\lower4pt\hbox{\hskip1pt$\sim$}}
    \raise1pt\hbox{$>$}}}         

\begin{document}

\vspace*{4cm}

\title{
\renewcommand{\thefootnote}{\fnsymbol{footnote}}
PARTON ENERGY LOSS IN COLLINEAR EXPANSION 
\footnote{Contributed to 43rd Rencontres de Moriond on 
QCD and High-Energy Hadronic Interactions, La Thuile, Italy, 
8-15 Mar 2008. The work is done in collaboration
with P.~Aurenche and H.~Zaraket}
\renewcommand{\thefootnote}{\arabic{footnote}}
}

\author{ 
B.G.~ZAKHAROV
}

\address{
L.D. Landau Institute for Theoretical Physics,
        GSP-1, 117940,\\ Kosygina Str. 2, 117334 Moscow, Russia\\
}

\maketitle\abstracts{
We demonstrate that the $N\!=\!1$ rescattering contribution 
to the gluon radiation from a fast massless quark 
in $eA$ DIS vanishes in the collinear approximation. 
It is shown that the nonzero 
$N\!=\!1$ gluon spectrum obtained in
the higher-twist approach by Guo, Wang and Zhang \cite{W1,W2}
is a consequence of unjustified neglecting some important
terms in the collinear expansion.
}

\noindent{\bf 1.}
There are several approaches to
the induced gluon emission from fast partons
due to multiple scattering in cold nuclear matter and hot quark-gluon
plasma \cite{BDMPS,GLV1,W1,W2,Z1}. The most general approach
to this phenomenon is the so-called light-cone path integral (LCPI)
approach \cite{Z1}
(for reviews, see \cite{Z_YAF,Z_S}).
This formalism 
reproduces the predictions of the BDMPS \cite{BDMPS} and GLV 
\cite{GLV1} approaches in their applicability regions \cite{BSZ,Wied}
(at strong Landau-Pomeranchuk-Migdal suppression for massless partons,
and thin plasmas, respectively).
However, the relation between the LCPI approach and 
the higher-twist formalism by Guo, Wang and Zhang (GWZ)
\cite{W1,W2} is not clear.
The GWZ approach
\cite{W1,W2} is based on the Feynman diagram formalism and 
collinear expansion.
It includes only the $N\!=\!1$ rescattering and has  
originally been derived for the gluon emission
from a fast quark produced in $eA$ DIS. 
The analyses \cite{BDMPS,GLV1,Z1} neglect the quantum nonlocality in 
production of fast partons.
In \cite{W1,W2}
the nonlocal fast quark production and gluon emission have been treated 
on even footing.
However, one can show that in the applicability region
of the GWZ formalism the quantum nonlocality in the quark production
is not important for gluon emission, and 
the gluon spectrum of \cite{W1,W2} 
should coincide with the $N\!=\!1$ gluon spectrum in the LCPI 
approach \cite{Z_OA}.
But this is not the case.
The GWZ gluon spectrum predicted in \cite{W1,W2} contains the logarithmically
dependent nucleon 
gluon density, which is absent in the LCPI calculations \cite{Z_OA}.

We will demonstrate that the approximations used in \cite{W1,W2} 
really lead to a disagreement with the LCPI
approach \cite{Z1}.
However, contrary to the results of \cite{W1,W2} the correct use of
the collinear expansion gives a zero gluon spectrum.
The nonzero spectrum obtained in \cite{W1,W2}
is a consequence of unjustified neglecting some important terms.

\noindent{\bf 2.}
We consider the gluon emission from a fast quark produced in $eA$ DIS
for the Bjorken variable $x_{B}$ and photon virtuality
$Q$. 
The transverse momentum integrated 
distribution for the $gq$ final state can be described in 
terms of the  semi-inclusive nuclear hadronic tensor 
$dW^{\mu\nu}_{A}/dz$
(hereafter $z=\omega/E$, 
where $\omega$ is the gluon energy and $E$ is the 
struck quark energy).
The spin effects in the rescatterings of 
fast partons can be neglected. This ensures that the spin structure of 
$dW^{\mu\nu}_{A}/dz$ is the same as for the usual hadronic tensor 
$W^{\mu\nu}_{N}$ in $eN$ DIS. 
It allows one to describe the gluon emission in terms of the scalar
semi-inclusive quark distribution.
Neglecting the EMC and shadowing
effects it can be written as 
\beq
\frac{df_{A}}{dz}=
\int d\rb n_{A}(\rb)
\frac{df_{N}(\rb)}{dz}\,,
\label{eq:30}
\eeq 
where $df_{N}(\rb)/dz$ is the in-medium semi-inclusive quark distribution
for a nucleon located at $\rb$, and $n_{A}(\rb)$ is the nucleus number density. 

In the LCPI approach \cite{Z1} the matrix element of the $q\rightarrow gq'$
in-medium transition is written in terms of the wave 
functions of the initial quark and final quark and gluon in the 
nucleus color field (we omit the color factors and indices)
\beq
\langle gq'|\hat{S}|q\rangle=ig\int\! dy 
\bar{\psi}_{q'}(y)\gamma^{\mu}A_{\mu}^{*}(y)\psi_{q}(y)\,.
\label{eq:40}
\eeq
Each quark wave function in (\ref{eq:40}) is written as 
$
\psi(y)=\frac{\exp(-ip^{-}y^{+})}{\sqrt{2p^{-}}}
\hat{u}_{\lambda}
\phi(y^{-},\vec{y}_{T})\,,$
where $\lambda$ is the quark helicity,
$\hat{u}_{\lambda}$ is the Dirac spinor operator,
$y^{\pm}=(y^{0}\pm y^{3})/\sqrt{2}$.
The $y^{-}$ dependence of the transverse wave functions
$\phi$ is governed by the two-dimensional 
Schr\"odinger equation 
\beq
i\frac{\partial\phi(y^{-},\vec{y}_{T})}{\partial
y^{-}}={\Big\{}\frac{
[(\vec{p}_{T}-g\vec{A}_{T})^{2}
+m^{2}_{q}]} {2\mu}
+gA^{+}{\Big\}}
\phi(y^{-},\vec{y}_{T})\,
\label{eq:60}
\eeq
with $\mu=p^{-}$. The wave function of the emitted
gluon can be represented in a similar way. 
The $y^{-}$ evolution of the transverse wave functions can be described 
in terms
of the Green's function 
${\cal{K}}$
for the Schr\"odinger equation (\ref{eq:60}).
One can show that for the gauges with potential vanishing at large
distances (say, covariant gauges, or Coulomb gauge) one can ignore
the transverse potential $\vec{A}_{T}$. 
To calculate the $N\!=\!1$ 
rescattering contribution we need not use the path integral
representation for the Green's functions (which is used at
the last stage of calculations \cite{Z1}).
To obtain the $N\!=\!1$ contribution it is enough to expand ${\cal{K}}$
to the second order in the external potential.
It allows one to describe the induced gluon emission in $eA$ DIS in terms 
of the free Green's functions $K$ for fast partons and the gluon correlator 
$\langle A^{+}(y_{1})A^{+}(y_{2})\rangle$ in the nucleus.
Diagrammatically it is represented by a set of 
diagrams like shown in 
Fig.~1 in which the horizontal solid line corresponds to $K$ ($\rightarrow$)
and $K^{*}$ ($\leftarrow$), the gluon line shows the 
gluon correlators,
the vertical dashed line shows the transverse density
matrices of the final quark and gluon at very large $y^{-}$.
\vspace{.15cm}
\begin{center}
\epsfig{file=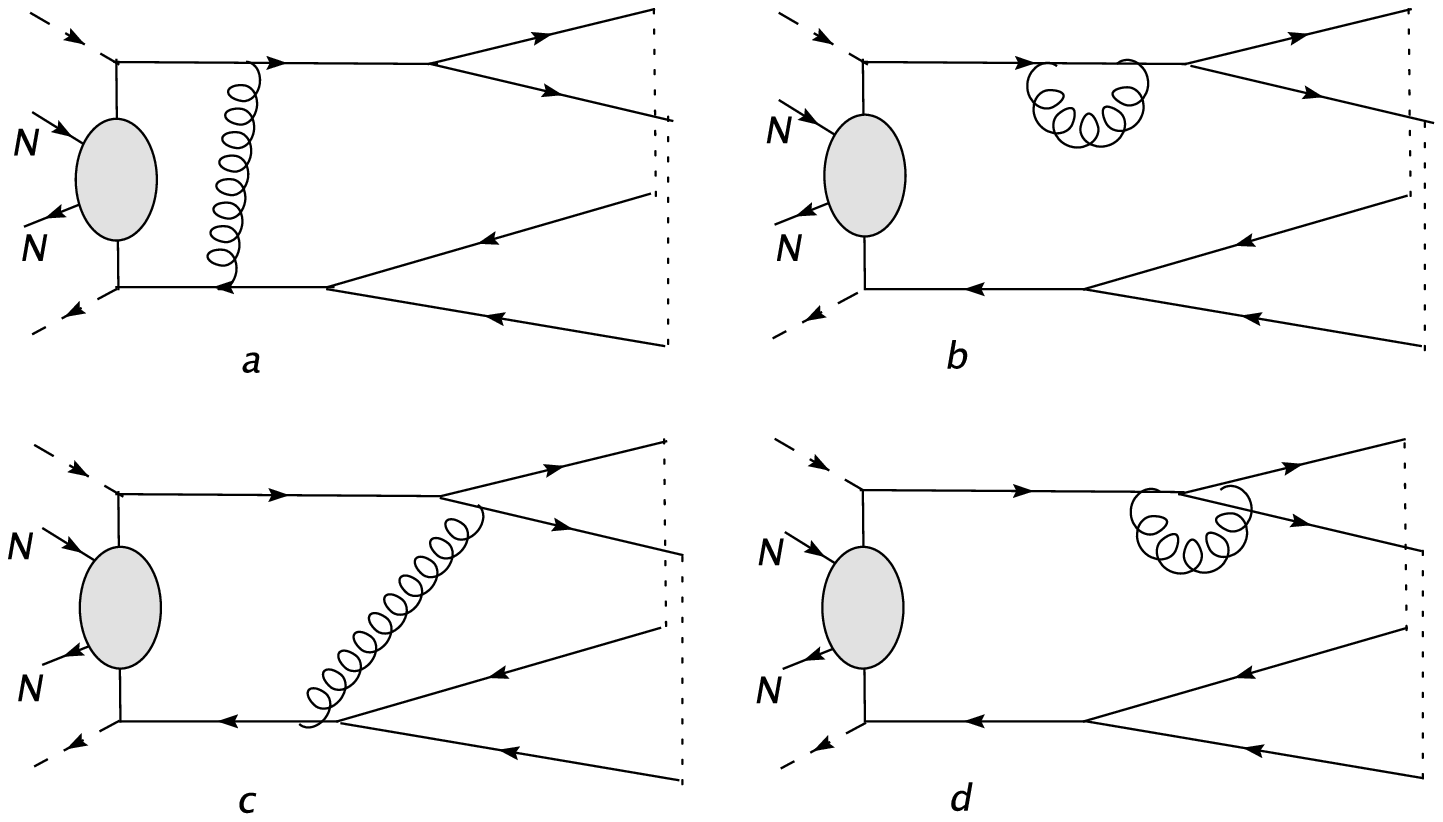,height=5.3cm,angle=0}
\end{center}
\begin{center}Figure 1. 
\end{center}
The typical difference in the coordinate $y^{-}$ for the 
upper and lower $\gamma^{*}qq$ vertices (which gives the scale of the 
quantum nonlocality of the fast quark production)
is given by the well known Ioffe
length $L_{I}=1/m_{N}x_{B}$. For the 
nucleon quark distribution $L_{I}$ is the dominating scale
in the Collins-Soper formula \cite{Soper} 
$f_{N}=\frac{1}{4\pi}\int dy^{-}e^{ix_{B}P^{+}y^{-}}
\langle N|\bar{\psi}(-y^{-}/2)\gamma^{+}\psi(y^{-}/2)|N\rangle$.
For the $gq$ final state the 
integration over the $y^{-}$ coordinate
of the $\gamma^{*}qq$ vertex is affected by the integration over the 
positions of rescatterings and the $q\rightarrow gq$ splitting.
However, for moderate $x_{B}$ when $L_{I}\ll R_{A}$ 
one can neglect the effect of rescatterings on the integration over $y^{-}$.
For production of the final $gq$ 
states with $M^{2}_{gq}\ll Q^{2}$ the restriction on $y^{-}$
from the splitting point can also be ignored.
Indeed, the typical scale in integrating over
the splitting points is given by the gluon formation length $L_{f}\sim
E/M^{2}_{gq}$ which is much bigger than $L_{I}$ at $M^{2}_{gq}\ll Q^{2}$.
This is valid for both the vacuum DGLAP and the 
induced gluon emission.
Also, at 
$L_{I}\ll L_{f}$
one can take for the lower limit of the integration over the 
splitting points for the upper and lower parts of the diagrams in Fig.~1
the position of the struck nucleon. 
Then the quark production and gluon emission become independent and 
the ${df_{N}(\rb)}/{dz}$ can be approximated 
by the factorized form 
\beq
\frac{df_{N}(\rb)}{dz}\approx f_{N}\frac{dP(\rb)}{dz}\,,
\label{eq:90}
\eeq 
where $dP/dz$ is the induced gluon spectrum 
described by the right parts of the diagrams evaluated neglecting
the quantum nonlocality of the fast quark production.

Due to confinement the typical separation of the arguments in the gluon 
correlators  
is of the order of the nucleon radius, $R_{N}$. 
It allows one to replace the fast parton propagators between 
the gluon fields in the graphs like Fig.~1b by $\delta$ 
functions in impact parameter space. This approximation is valid 
for parton energy $\gg 1/R_{N}$.
It follows from the Schr\"odinger diffusion
relation for the parton transverse motion $\rho^{2}\sim L/E$.
Also, the smallness of the fast parton diffusion radius at the 
longitudinal scale $\sim R_{N}$ allows one
to replace in other transverse Green's functions the $y^{-}$
coordinates by the mean values of the arguments of the vector potentials
in the gluon correlators. 
This approximation corresponds to a picture with 
rescatterings of fast partons on zero thickness scattering centers (nucleons).
The inequality $\omega \gg 1/R_{N}$ for the emitted
gluon is equivalent to $R_{N}\ll L_{f}$.
For this reason, in the picture of thin nucleons
the contribution of the graphs like Fig.~1c,d with gluon correlators 
connecting the initial quark and final quark or gluon 
can be neglected since they are suppressed by the small factor
$R_{N}/L_{f}$. 
These approximations have been used in the original formulation of
the LCPI approach \cite{Z1} (the BDMPS \cite {BDMPS} and GLV 
\cite{GLV1} approaches use them as well).
Note that (similarly to the case of the quark-gluon plasma \cite{AZ}) 
each gluon correlator 
appears only in the form of an integral over $\Delta y^{-}=y^{-}_{2}-y^{-}_{1}$
and at $y^{+}_{2}=y^{+}_{1}$. One can easily show that 
this ensures  gauge invariance 
of the result (to leading order in $\alpha_{s}$).

In \cite{W1,W2} the gluon emission in $eA$ DIS is described by 
the diagrams like shown in Fig.~2. 
\begin{center}
\epsfig{file=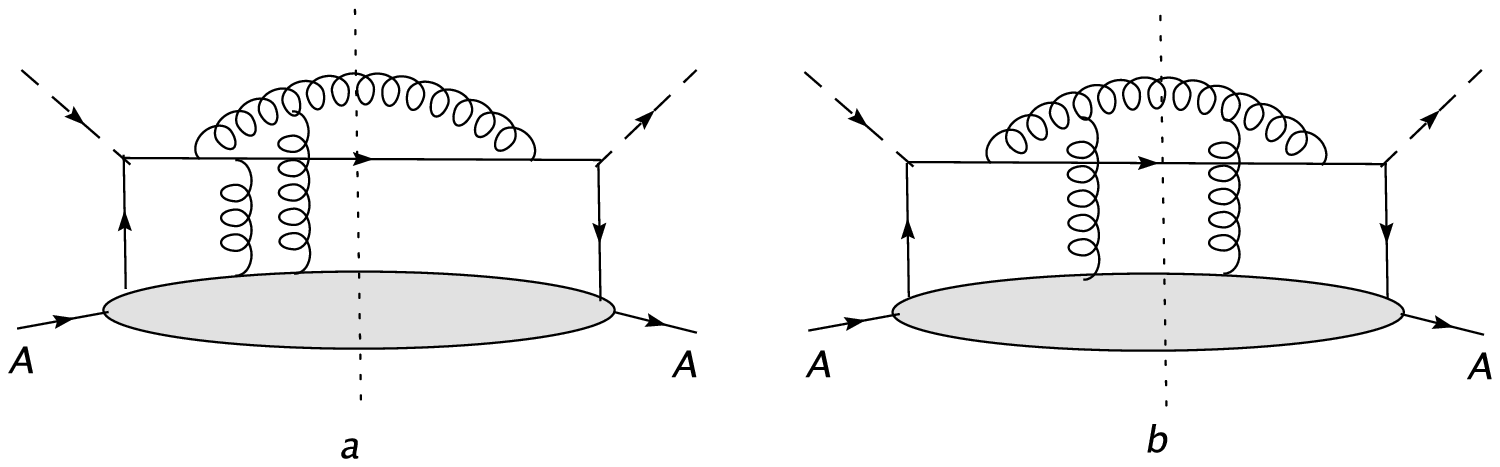,height=3.7cm,angle=0}
\end{center}
\begin{center}Figure 2. 
\end{center}
The lower soft part is expressed in terms of the matrix element
$
\langle A|\bar{\psi}(0)A^{+}(y_{1})A^{+}(y_{2})\psi(y_{3})|A\rangle
$, and  the upper hard parts are calculated perturbatively. 
Due to conservation of the large $p^{-}$ momenta of fast partons 
in the Feynman propagators only the Fourier components with 
$p^{-}>0$ are important. It means that the Feynman propagators are 
effectively reduced to the retarded (in $y^{-}$ coordinate) ones. 
One can show that the Feynman diagram treatment of \cite{W1,W2}
is equivalent to that in terms of the transverse Green's functions. 
Indeed, using 
the representation
\beq
K(\vec{y}_{T,2},y^{-}_{2}|\vec{y}_{T,1},y^{-}_{1})=
i\int \frac{dp^{+}d\vec{p}_{T}}{(2\pi)^{3}}
\frac
{\exp{[
-ip^{+}(y^{-}_{2}-y{-}_{1})+i\vec{p}_{T}(\vec{y}_{T,2}-\vec{y}_{T,1})
]}}{p^{+}-
\frac{\vec{p}_{T}^{\,2}+m^{2}}{2p^{-}}+i0}
\label{eq:110}
\eeq
one can write the retarded quark propagator as
\bea
G_{r}(y_{2}-y_{1})=\frac{1}{4\pi}
\int_{0}^{\infty}\frac{dp^{-}}{p^{-}}
e^{-ip^{-}(y^{+}_{2}-y^{+}_{1})}\left[
\sum\limits_{\lambda}\hat{u}_{\lambda}\bar{\hat{u}}_{\lambda}
K(\vec{y}_{T,2},y^{-}_{2}|\vec{y}_{T,1},y^{-}_{1})
\right.
\nonumber\\
\left.
+i\gamma^{-}\delta(y^{-}_{2}-y^{-}_{1})
\delta(\vec{y}_{T,2}-\vec{y}_{T,1})\right]
\,.
\label{eq:120}
\eea
Here $\hat{u}_{\lambda}$ and $\bar{\hat{u}}_{\lambda}$ act on the variables
with indices 2 and 1, respectively. The last term in (\ref{eq:120}) 
is the so-called contact term. It does not propagate
in $y^{-}$ and can be omitted in calculating the 
nuclear final-state interaction effects for fast partons. 
Using (\ref{eq:120}) and a similar representation for the gluon 
propagator the hard parts in the higher-twist method can be 
represented in terms
of the transverse Green's functions as it is done in the LCPI treatment.
We emphasize that the description of the hard parts in terms of the transverse
Green's functions automatically includes all the processes in the 
GWZ approach (hard-soft,
double-hard, and interferences in the terminology of \cite{W1,W2}).

\vspace{.1cm}
\noindent{\bf 3.}
The calculation of the diagrams like shown in Fig.~1 
is simplified by noting that the free transverse Green's function 
can be written as
\beq
K(\vec{y}_{T,2},y^{-}_{2}|\vec{y}_{T,1},y^{-}_{1})=
\theta(y^{-}_{2}-y^{-}_{1})\sum_{p_{T}}
\phi_{p_{T}}(\vec{y}_{T,2},y^{-}_{2})
\phi_{p_{T}}^{*}(\vec{y}_{T,1},y^{-}_{1})\,,
\label{eq:130}
\eeq
where $\phi_{p_{T}}(\vec{y}_{T},y^{-})$
 is the plane wave solution to the
Schr\"odinger equation for $A^{\mu}=0$ with the transverse momentum 
$\vec{p}_{T}$. It allows one to represent the upper and lower parts of
the diagrams shown in Fig.~1 in the form
$
\int d y^{-}d\vec{y}_{T}
 \phi_{q'}^{*}(\vec{y}_{T},y^{-})
\phi_{g}^{*}(\vec{y}_{T},y^{-})
\phi_{q}(\vec{y}_{T},y^{-})
$
where the outgoing and incoming wave functions have the form of the plane
waves with sharp changes of the transverse momenta at the points of 
interactions with the external gluon fields.
This method has previously been used in \cite{Z_kin} for investigation
of the role of the finite kinematical boundaries. 
All the hard parts evaluated with the help of the 
plane waves agree with that obtained in Refs. \cite{W1,W2}.

The sum of the complete set of the diagrams contributing to the $N\!=\!1$
spectrum can be written as \cite{Z_OA,Z_kin}
\beq
\frac{dP(r)}{dz}=\int\limits_{r_{3}}^{\infty} d\xi 
n_{A}(\vec{r}_{T},r_{3}+\xi)\frac{d\sigma(z,\xi)}{dz}\,.
\label{eq:140}
\eeq
Here $d\sigma(z,\xi)/dz$ is the cross section of gluon emission
from the fast quark produced at distance $\xi$ from the scattering nucleon.
At $z\ll 1$ (we consider the soft gluon emission
just to simplify the formulas) for massless partons it reads \cite{Z_OA,Z_kin} 
\beq
\frac{d\sigma (z,\xi)}{dz}=
\frac{2\alpha_{s}^{2}[1+(1-z)^{2}]}{z}
\int 
d\vec{p}_{T}
\frac{d\vec{k}_{T}}{\vec{k}_{T}^{\,2}}  
\frac{xdG(k^{2}_{T},x)}{d\vec{k}_{T}}
{H(\vec{p}_{T},\vec{k}_{T},z,\xi) }\,,
\label{eq:150}
\eeq
\beq
H(\vec{p}_{T},\vec{k}_{T},z,\xi)\!=\!
\left[\frac{1}
{\vec{p}_{T}^{\,2}}-
\frac{(\vec{p}_{T}-\vec{k}_{T})\vec{p}_{T} }
{\vec{p}_{T}^{\,\,2}
(\vec{p}_{T}-\vec{k}_{T})^{2}
}\right]
\cdot
\left[1-
\cos\left(\frac{i\vec{p}_{T}^{\,2}\xi}{2Ez(1-z)}
\right)
\right]\,.
\label{eq:160}
\eeq
Here the limit $x\rightarrow 0$ is implicit, 
$dG(k^{2}_{T},x)/d\vec{k}_{T}$ is the unintegrated nucleon gluon density
in the Collins-Soper form \cite{Soper},
which 
at $x\ll 1$ can also be written as
\beq
\frac{dG(k^{2}_{T},x)}{d\vec{k}_{T}}=
\frac{N_{c}^{2}-1}{x32\pi^{4}\alpha_{s}C_{F}}
\int  d\vec{\rho}\exp{(-i\vec{k}_{T}\vec{\rho})}
{\nabla}^{2}\sigma(\rho)\,,
\label{eq:180}
\eeq
where $\sigma(\rho)$ is the well known dipole cross section.

The collinear expansion corresponds to replacement of the hard part $H$ by
its second order expansion in $\vec{k}_{T}$ (we suppress all the arguments
except for $\vec{k}_{T}$ for clarity)
\beq
H(\vec{k}_{T})\approx
H(\vec{k}_{T}=0)+
\left.\frac{\partial H}
{\partial k^{\alpha}_{T}}\right|_{\vec{k}_{T}=0}k^{\alpha}_{T}
+
\left.\frac{\partial^{2} H}
{\partial k^{\alpha}_{T} \partial k^{\beta}_{T}}
\right|_{\vec{k}_{T}=0}\cdot
\frac{k^{\alpha}_{T}k^{\beta}_{T}}{2}
\,.
\label{eq:200}
\eeq
Only the second order term in (\ref{eq:200}) is important, which
gives to a logarithmic accuracy 
$
d\sigma (z,\xi)/dz
\propto 
\int d\vec{p}_{T} 
x G(p^{2}_{T},x)
\nabla^{2}_{k_{T}}H(\vec{p}_{T},\vec{k}_{T},z,\xi)
|_{\vec{k}_{T}=0} 
$.
But from (\ref{eq:160}) one can easily obtain 
$\nabla_{k_{T}}^{2}H|_{\vec{k}_{T}=0}\!=0\!$.
It is also seen from averaging of the hard part over
the azimuthal angle of $\vec{k}_{T}$ which gives 
$\langle H(\vec{p}_{T},\vec{k}_{T},z,\xi)\rangle
\propto \theta(k_{T}-p_{T})$.
Thus, contrary to the expected 
dominance of the region $k_{T}\lsim p_{T}$ only the 
region $k_{T}>p_{T}$ contributes to the gluon emission, and formal use of
the collinear expansion gives completely wrong result with zero gluon spectrum.

The zero $N\!=\!1$ gluon spectrum in the collinear
approximation agrees with prediction of the harmonic oscillator
approximation in the BDMPS \cite{BDMPS} and
LCPI \cite{Z1} approaches. 
The oscillator approximation in \cite{BDMPS,Z1} corresponds to
the quadratic parametrization of the dipole cross section
$\sigma(\rho) = C\, \rho^2$. This parametrization is equivalent to
the approximation of the vector potential by the linear expansion
$
A^{+}(y^{-},\vec{y}_{T}+\vec{\rho})\approx
A^{+}(y^{-},\vec{y}_{T})+
\vec{\rho}
\nabla_{y_{T}}
A^{+}(y^{-},\vec{y}_{T})\,
$
which can be traced to the
collinear expansion in momentum space.
The first term in the expansion in the density of the spectrum in 
the oscillator approximation in the BDMPS and LCPI approaches 
corresponds to  $N\!=\!2$, and the term with $N\!=\!1$ rescattering
is absent \cite{Z_OA}. 
In terms of the representation (\ref{eq:150}) absence of the $N\!=\!1$ 
contribution in the oscillator approximation is a consequence of 
the fact that in this case 
$dG/d\vec{k}_{T}\propto \delta(\vec{k}_{T})$ (as one sees from (\ref{eq:180})).

\vspace{.1cm}
\noindent{\bf 4.}
The vanishing $N\!=\!1$ spectrum is in a clear contradiction with
the nonzero result of \cite{W1,W2}. This discrepancy is strange enough 
since Eqs. (\ref{eq:140})-(\ref{eq:160}) are completely equivalent to the
formulation of \cite{W1,W2} in the approximation of thin nucleons,
and the effects beyond this approximation cannot be evaluated in the formalism
\cite{W1,W2}. This puzzle has a simple solution.
In \cite{W1,W2} the nonzero second derivative
of the hard part  comes from the graph shown in Fig.~2b (at $z\ll 1$).
The authors use for the integration variable in the hard part of
this graph the transverse momentum of the final gluon, $\vec{l}_{T}$.
The $\vec{l}_{T}$-integrated
hard part obtained in \cite{W2} (Eq. 15 of \cite{W2}) reads
(up to an unimportant factor)  
\beq
H(\vec{k}_{T})\propto \int \frac{d\vec{l}_{T}}
{(\vec{l}_{T}-\vec{k}_{T})^{2}} R(y^{-},y^{-}_{1},y^{-}_{2},
\vec{l}_{T},\vec{k}_{T})\,,
\label{eq:260}
\eeq
where 
\bea
{R}(y^{-},y^{-}_{1},y^{-}_{2},
\vec{l}_{T},\vec{k}_{T})=\frac{1}{2}
\exp{
\left[i
\frac
{y^{-}(\vec{l}_{T}-\vec{k}_{T})^{2}-
(1-z)(y^{-}_{1}-y^{-}_{2})(\vec{k}_{T}^{\,2}
-2\vec{l}_{T}\vec{k}_{T})}{2q^{-}z(1-z)}
\right]
}\nonumber\\
\times\left[
1-\exp{
\left(i
\frac{
(y^{-}_{1}-y^{-})(\vec{l}_{T}-\vec{k}_{T})^{2}}
{2q^{-}z(1-z)}\right)}
\right]
\cdot
\left[
1-\exp{
\left(-i
\frac{
y^{-}_{2}(\vec{l}_{T}-\vec{k}_{T})^{2}}
{2q^{-}z(1-z)}\right)}
\right]
\label{eq:270}
\eea
is an analog of the last factor in the square brackets 
in (\ref{eq:160}) 
for $y^{-}\ne 0$, $y^{-}_{1}\ne y^{-}_{2}$
($y^{-}$, $y^{-}_{1,2}$ correspond to the quark interactions 
with the virtual photon and $t$-channel gluons,
our $z$ equals $1-z$ in \cite{W1,W2}).
In calculating 
$\nabla_{k_{T}}^{2}H(\vec{k}_{T})$
the authors differentiate only the factor
$1/(\vec{l}_{T}-\vec{k}_{T})^{2}$. However, the omitted terms 
from the factor $R$
are important. After the $\vec{l}_{T}$ integration they almost completely 
cancel the contribution
from the $1/(\vec{l}_{T}-\vec{k}_{T})^{2}$ term. Indeed, after 
putting $y^{-}_{1}=y^{-}_{2}$ and changing the 
variable 
$ \vec{l}_{T}\rightarrow (\vec{l}_{T}+\vec{k}_{T})$ 
the right-hand part of (\ref{eq:260}) does not depend on
$\vec{k}_{T}$ at all. We emphasize that even without the change of the 
variable (\ref{eq:260}) leads to $\nabla_{k_{T}}^{2}H|_{\vec{k}_{T}=0}\!=0\,$
if one performs differentiating correctly which is not done in \cite{W1,W2}. 
The difference between  $y^{-}_{1}$ and $y^{-}_{2}$
gives some nonzero contribution to the
$\nabla_{k_{T}}^{2}H|_{\vec{k}_{T}=0}$  
suppressed by the small factor $\sim \! (R_{N}/L_{f})^{2}$. 
Such contributions may be viewed as zero since they are 
beyond predictive accuracy of 
the approximations used in \cite{W1,W2}
\footnote[1]{
Keeping the correction from the nonzero $y^{-}$ in (\ref{eq:270})
also does not make sense under the approximations used in \cite{W1,W2}.}.

\vspace{.1cm}
\noindent {\bf 5}. 
In summary, we have demonstrated that 
the collinear expansion
fails in the case of gluon emission from a fast massless quark produced
in $eA$ DIS. In this approximation the $N\!=\!1$ rescattering contribution 
to the gluon spectrum vanishes.
The nonzero gluon spectrum obtained in \cite{W1,W2}
is a consequence of unjustified 
neglecting some important terms
in the collinear expansion.
The established facts demonstrate that the GWZ approach \cite{W1,W2}
is wrong. Its predictions 
for $eA$ DIS and jet quenching in $AA$ collisions 
do not make sense.

\section*{Acknowledgments}
This work is supported in part by the grant RFBR
06-02-16078-a and 
the LEA Physique Th\'eorique de la
Mati\'ere Condes\'ee.

\end{document}